\documentclass[aps,pre,twocolumn,superscriptaddress]{revtex4-1}
\usepackage{mathtools,amssymb}
\usepackage{graphicx}
\usepackage{epstopdf}
\usepackage{color}
\usepackage{xr}

\begin{document}
\title{Nonlinear refrigerator with a finite-sized cold heat bath}
\author{I. Iyyappan}
\email[email: ]{iyyap.si@gmail.com}
\affiliation{The Institute of Mathematical Sciences,\\ CIT Campus, Taramani, Chennai 600113, India.} 
\affiliation{Homi Bhabha National Institute, Training School Complex,\\ Anushakti Nagar, Mumbai 400094, India.}
\affiliation{Division of Sciences, Krea University, Sri City 517646, India.}

\author{Yuki Izumida}
\email[email: ]{izumida@k.u-tokyo.ac.jp}
\affiliation{Department of Complexity Science and Engineering, Graduate School of Frontier Sciences,\\
The University of Tokyo, Kashiwa 277-8561, Japan.}
\author{Sibasish Ghosh}
\email[email: ]{sibasis@imsc.res.in}
\affiliation{The Institute of Mathematical Sciences,\\ CIT Campus, Taramani, Chennai 600113, India.} 
\affiliation{Homi Bhabha National Institute, Training School Complex,\\ Anushakti Nagar, Mumbai 400094, India.}

\begin{abstract}
We study the refrigerator working between a finite-sized cold heat bath and an infinite-sized hot heat bath (environment) in the nonlinear response regime. We assume that the initial temperature $T_i$ of the finite-sized cold heat bath satisfies $T_i\leq T_h$, where $T_h$ is the temperature of the hot heat bath. By consuming the input power, the refrigerator transfers the heat from a finite-sized cold heat bath to the hot heat bath. Hence, the temperature of the finite-sized cold heat bath decreases until it reaches the desired low-temperature $T_f$. By minimizing the input work for the heat transport process, we derive the optimal path for temperature change. We calculate the coefficient of performance as a function of average input power. We also obtain the bounds for the coefficient of performance by applying the asymmetric dissipation limits. For the parameter values considered in this study, we observe that the relation between the coefficient of performance and the average input power strongly depends on the nature of the finite-sized heat cold bath.
\end{abstract}
\maketitle

\section{Introduction}
Optimizing the performance of heat engines and refrigerators is a problem that has been studied for centuries. Efficiently utilizing limited energy resources is vital to fulfilling the power demands of our present world. The first optimization of energy conversion from heat to work goes back to 1824 by Sadi Carnot \cite{carn}. He showed that the efficiency of a heat engine attains its maximum when it works reversibly, and it is given by the Carnot value $\eta_{\rm C}=1-T_c/T_h$. Here, $T_c$ and $T_h$ are the temperatures of the cold and hot heat baths, respectively. The inverse operation of a heat engine is called the refrigerator which transports heat from the cold heat bath to the hot heat bath by consuming the input work. The maximum coefficient of performance (COP) of the refrigerator is given by $\epsilon_{\rm C}=T_c/(T_h-T_c)$ \cite{cal}. 

Carnot devices are impractical. In the case of heat engines, it delivers zero power output, and for refrigerators, it provides zero cooling power. Almost half a century ago in 1975, Curzon and Ahlborn constructed a finite-time Carnot heat engine (endorevesible heat engine) and showed the efficiency at maximum power as $\eta_{\rm CA}=1-\sqrt{T_c/T_h}$ \cite{cur22}. This simple and elegant result opened up a new field of study called finite-time thermodynamics \cite{and208,and157,rub127,sal211,dev570,sal354}. Later, the COP of endorevesible Carnot refrigerators at maximum cooling power was studied by Agrawal and Menon \cite{agr531}.

In 2005, Van den Broeck investigated the generic heat engine using linear irreversible thermodynamics (LIT) without invoking any specific heat transfer law to calculate the heat injection and rejection \cite{van190}. Subsequently, de Cisneros \textit{et al.} studied the performance of linear irreversible refrigerators \cite{jim057}. For refrigerators, the LIT framework can be described as follows. When the temperature difference between the hot and cold heat baths, $\Delta T=T_h-T_c$ is very small compared to the reference values (i.e., $T_h\approx T_c$), and the external mechanical force $F$ applied to the refrigerator is also very small, then, the Onsager relations can be written as \cite{jim057,ons226}
\begin{equation}\label{a}
J_1=L_{11}X_1+L_{12}X_2,
\end{equation}
\begin{equation}\label{b}
J_2=L_{21}X_1+L_{22}X_2.
\end{equation}
Here, $L_{ij}$ ($i, j=1, 2$) is the Onsager coefficient. We defined the mechanical thermodynamic force $X_1=F/T_h$, and the corresponding flux (e.g., velocity of the piston),  $J_1=\dot{x}$. We also defined the thermal thermodynamic force $X_2=1/T_h-1/T_c$, and the corresponding heat flux $J_2=\dot{Q_c}$, which is absorbed from the cold heat bath at temperature $T_c$. The input power is given by 
\begin{equation}\label{b2}
P=F\dot{x}=T_hJ_1X_1.
\end{equation}
The second law of thermodynamics imposes a constraint on the Onsager coefficients,
\begin{equation}
L_{11}\geq0,\; L_{22}\geq0,\; L_{11}L_{22}-L_{12}L_{21}\geq0.
\end{equation}
In our study, we assume the reciprocity relation, $L_{12}=L_{21}$ \cite{cal}. In real heat devices, they operate at a finite rate, and hence heat dissipation is unavoidable. To include the heat dissipation beyond the framework of LIT, Izumida, and Okuda added the nonlinear term to the heat flux, $J_2$ \cite{izu100}. It is named a minimally nonlinear irreversible model. One can show that the minimally nonlinear irreversible model includes the low-dissipation model as a special case \cite{izu100,esp150}. The minimally nonlinear irreversible heat devices were widely studied in the literature \cite{izu1000,she012,lon062,she022,lon052,izu052,lon0521,iyy012,liu717}.

In addition to the finite-time constraint, practical heat devices have finite-sized energy source or sink, which have significant consequences on their operation and performance \cite{ond681,ond472,and270,and266,lef701,yan811}. Izumida and Okuda investigated the heat engine working between the finite-sized hot heat source and an infinite-sized cold heat bath using the linear irreversible thermodynamics \cite{izu180}. Inspired by their study, several authors focused on finding the finite-sized bath effects on the efficiency of heat engines \cite{wan062,wan012,joh100,joh012,iyy500,ma100,yua022,hua125}. 

Refrigerators were used in several places. For example, food storage at home, industrial manufacturing, research labs, space stations, medical fields, etc. Whenever we want to cool the system, it always has a finite size. For example, consider air inside the fridge or the living room, which acts as a cold bath. It contains only a finite number of molecules and hence its heat capacity becomes finite. The absorbed heat from the cold bath is rejected to the environment, which is extremely large compared to the cold bath size. Therefore, we consider the environment as a hot heat bath with infinite heat capacity. Despite its practical significance, we found little study on refrigerators working with a finite-sized cold heat bath and an infinite-sized hot heat bath except for \cite{ron040}. Therefore, in this work, we investigate the nonlinear irreversible refrigerator which includes the dissipation effect working with a finite-sized cold heat bath and an infinite-sized hot heat bath. We assume that initially the temperature of the finite-sized cold heat bath satisfies $T_i\leq T_h$, which we may encounter in our everyday life. A refrigerator starts transporting heat from the finite-sized cold heat bath to a hot heat bath by consuming the input work/power. We assume that the lowest temperature the finite-sized cold heat bath can reach is $T_f<T_i$. We optimize the input work required to reach the desired low-temperature $T_f$, and study the performance of the refrigerator.

This paper is organized as follows. In Sec.~\ref{Model}, we briefly describe the nonlinear irreversible refrigerator model, and in Sec.~\ref{Cold bath}, we apply the finite-sized cold heat bath effect to this model. In Sec.~\ref{Optimization}, we introduce the optimization method, and consider the analytically feasible case to which we can formulate the finite-time COP.
In Sec.~\ref{Example}, we model the three different finite-sized cold heat baths as the demonstration of the theory. Finally, we conclude in Sec.~\ref{Conclusion}.

\begin{figure}[t!]
	\centering
	\includegraphics[scale=0.5,angle=0]{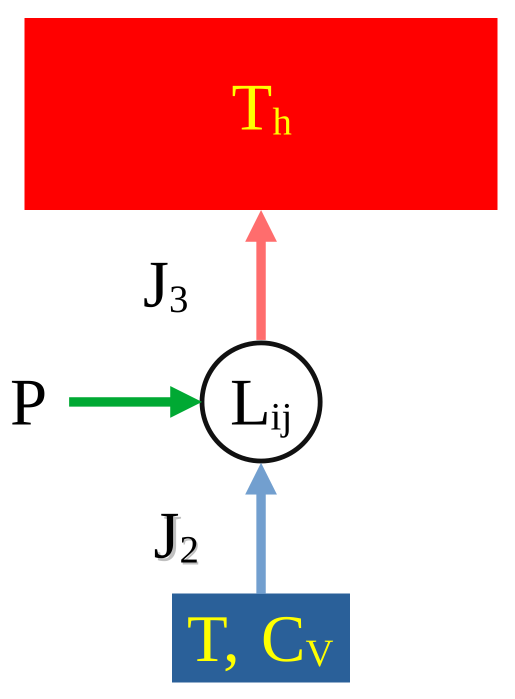}
	\caption{\label{fig:sche} The schematic diagram of a minimally nonlinear irreversible refrigerator working with a finite-sized cold heat bath and an infinite-sized hot heat bath.}
\end{figure}

\section{Non-linear Refrigerator}\label{Model}
The schematic diagram of a minimally nonlinear irreversible refrigerator is given in Fig.~\ref{fig:sche}, where we denote the temperature of the finite-sized cold heat bath by $T$ instead of $T_c$, and $C_V$ is the heat capacity at constant volume. In our work, we assume that the volume of the cold heat bath does not change during the heat transfer process.

Izumida and Okuda explicitly added a second-order term in the heat flux in Eq.~(\ref{b}). Therefore, the extended Onsager relations are given by~\cite{izu100}
\begin{equation}\label{c}
J_1=L_{11}X_1+L_{12}X_2,
\end{equation}
\begin{equation}\label{d}
J_2=L_{21}X_1+L_{22}X_2-\gamma_c J_1^{2}.
\end{equation}
Here, $\gamma_c>0$ represents the strength of the dissipation to the cold heat bath. Solving Eq.~(\ref{c}) for $X_1$, and substituting it into Eq.~(\ref{d}), we get the input heat flux $J_2$ in terms of $J_1$ as 
\begin{equation}\label{e}
J_2=\frac{L_{21}}{L_{11}}J_1+L_{22}(1-q^{2})X_2-\gamma_c J_1^{2}.
\end{equation}
Here, $q\equiv L_{12}/\sqrt{L_{11}L_{22}}$ is the coupling strength parameter with $-1\leq q\leq 1$~\cite{ked189}. Using Eqs. (\ref{b2}), and (\ref{c}), we get the input power as
\begin{equation}\label{f}
P=\frac{L_{21}}{L_{11}} \left(\frac{T_h}{T}-1 \right) J_1+\frac{T_h}{L_{11}} J_1^{2}.
\end{equation}
Utilizing the first law of thermodynamics, we can calculate the rejected heat flux to the hot heat bath, $J_3=\dot{Q_c}+P$: 
\begin{equation}\label{g}
J_3=\frac{L_{21}}{L_{11}}\frac{T_h}{T}J_1+L_{22}(1-q^{2})X_2+\gamma_h J_1^{2},
\end{equation}
where $\gamma_h>0$ is the strength of the dissipation into the hot heat bath 
\begin{equation}\label{h}
\gamma_h \equiv \frac{T_h}{L_{11}}-\gamma_c.
\end{equation}
The positive constraint on $\gamma_c$, and $\gamma_h$ ensures the non-negativity of the entropy production rate for the heat transport process (see Eq. (16) of Ref. \cite{izu1000}). One can find that, the heat-leakage $L_{22}(1-q^2)X_2$ present in $J_2$, and $J_3$ does not affect the input power $P$ (see Eq. (\ref{f})) (Fig.~\ref{fig:sche}). Next, we apply the finite-sized cold heat bath effect to the nonlinear irreversible model.

\section{Finite-sized cold heat bath}\label{Cold bath}
Because we are assuming that the cold heat bath has finite size, its temperature $T$ is changeable in time (Fig.~\ref{fig:sche}). The refrigerator absorbs the heat flux from the finite-sized cold heat bath, and its initial temperature is $T_i\leq T_h$. When heat is absorbed from a cold heat bath, its temperature starts to decrease from $T_i$ until it reaches final temperature $T_f<T_i$ and we assume it takes time $\tau$. The input heat flux $J_2$ can be written in terms of the rate of change of temperature of the finite-sized cold heat bath as \cite{izu180}
\begin{equation}\label{i}
J_2=-C_V\dot{T}.
\end{equation}
By using Eqs.~(\ref{i}) and (\ref{e}), we get the flux $J_1$ in Eq.~(\ref{c}) in terms of $\dot{T}$ as
\begin{equation}\label{j}
J_1=\frac{L_{21}}{2L_{11}\gamma_c}\left(1\pm\sqrt{1+\frac{4L_{11}\gamma_cC_V}{L_{22}}\dot{T}}\right).
\end{equation}
Here, we have used the tight-coupling condition, that is $|q|=1$ ~\cite{ked189}. The tight-coupling condition $|q|=1$ means that the two fluxes are highly dependent or tightly coupled. Meanwhile, $q=0$ means that the two fluxes are completely independent.  Substituting Eq.~(\ref{j}) into Eq.~(\ref{g}) and using Eq.~(\ref{h}), we get the rejected heat flux
\begin{widetext}
\begin{equation}\label{k1}
J_3=L_{22}\gamma\left(1\pm\sqrt{1+\frac{4T_{h}C_V\dot{T}}{L_{22}\gamma}}\,\right)\left[\frac{\gamma-1}{4T_h}\left( 1\pm\sqrt{1+\frac{4T_{h}C_V\dot{T}}{L_{22}\gamma}}\right)+\frac{1}{2T}\right],
\end{equation}
\end{widetext}
where we defined
\begin{equation}
\gamma \equiv 1+\frac{\gamma_h}{\gamma_c}.
\end{equation}
Note that in the asymmetric dissipation limits, when $\gamma_h\rightarrow 0$, we have $\gamma\rightarrow 1$, and when $\gamma_c\rightarrow 0$, we have $\gamma\rightarrow \infty$. The total work consumed by the refrigerator is given by
\begin{equation}
W (\tau)=\int_{0}^{\tau}J_3 (t) dt-\int_{0}^{\tau}J_2 (t) dt.\label{eq.W_def}
\end{equation}
Using Eq.~(\ref{i}), we can evaluate the second integration of the right-hand side of Eq.~(\ref{eq.W_def}), which is the total heat absorbed from the finite-sized cold heat bath:
\begin{equation}\label{s}
Q_c=\int_0^\tau J_2 (t) dt=-\int_{T_i}^{T_f} C_VdT=-\Delta U,
\end{equation}
where $\Delta U$ is the internal energy change of the cold heat bath.

\section{Optimization and finite-time COP}\label{Optimization}
To minimize the input work given in Eq.~(\ref{eq.W_def}), we need to optimize the first integration of the right-hand side of Eq.~(\ref{eq.W_def}) using the Euler-Lagrange (E-L) method. The E-L equation for $J_3$ is
\begin{equation}\label{m}
\frac{d}{dt}\left(\frac{\partial J_3}{\partial \dot{T}}\right)-\frac{\partial J_3}{\partial T}=0.
\end{equation}
It is not possible to solve analytically the E-L equation for $J_3$ given in Eq.~(\ref{k1}) (for both the sign of the root). Therefore, 
we focus on near the quasistatic regime $J_1\approx 0$ as similar to the low-dissipation model~\cite{esp150}.
This amounts to simplify the situation by considering the rate of change of temperature $|\dot{T}|$ to be sufficiently small.
Hence, we have 
\begin{equation}\label{eq.approx_cond}
\Biggl|\frac{4T_{h}C_V\dot{T}}{L_{22}\gamma}\Biggr|\ll 1
\end{equation}
at any instant of the process. This allows us to expand the square root term only up to the second order in $\dot{T}$ in Eq.~(\ref{k1}). Moreover, it is sufficient to take the minus sign in Eq.~(\ref{k1}) for optimization (or in Eq.~(\ref{j})) (see Appendix~\ref{appendix_root1} for justification).
 
\subsection{Optimized finite-time COP}
For the negative root case in Eq.~(\ref{k1}), 
we get the approximated rejected heat flux as
\begin{equation}\label{w}
J_3\approx-C_VT_h\left[\frac{\dot{T}}{T}+\left(1-\gamma-\frac{T_h}{T}\right)\frac{C_V\dot{T}^{2}}{ L_{22}\gamma}\right].
\end{equation}
The E-L equation for the above $J_3$ becomes
\begin{equation}\label{z}
\left(\frac{T_h}{T}-1+\gamma\right)\frac{2C_V^{2}\ddot{T}}{L_{22}\gamma}+\frac{\partial}{\partial T}\left[\left(\frac{T_h}{T}-1+\gamma\right)\frac{C_V^{2}}{ L_{22}\gamma}\right]\dot{T}^{2}=0,
\end{equation}
Multiplying Eq. (\ref{z}) with $\dot{T}$ and after doing some mathematical manipulation, one can get
\begin{equation}\label{1a}
\frac{d}{dt}\left[\left(\frac{T_h}{T}-1+\gamma\right)\frac{C_V^{2}\dot{T}^{2}}{L_{22}\gamma}\right]=0.
\end{equation}
Integrating Eq.~(\ref{1a}), we get the following expression
\begin{equation}\label{1b}
\sqrt{\frac{T_h}{T}-1+\gamma}\frac{C_V\dot{T}}{\sqrt{L_{22}\gamma}}=\mathcal A,
\end{equation}
where $\mathcal A$ is an integration constant. Once again, integrating both sides of Eq. (\ref{1b}), we get
\begin{equation}\label{1c}
\mathcal A=\frac{\mathcal B}{\tau};\;\;\; \mathcal B\equiv \int_{T_i}^{T_f}\sqrt{\frac{T_h}{T}-1+\gamma} \frac{C_V}{\sqrt{L_{22}\gamma}}dT.
\end{equation}
By solving Eq.~(\ref{1b}), we may get an analytical solution of $T(t)$. See Appendix~\ref{comparison} for comparison between numerical and analytical solutions of an illustrative case.

Using Eqs.~(\ref{w}) and (\ref{1c}), we can calculate the total heat rejected to the hot heat bath as
\begin{equation}\label{1d}
Q_h=\int_0^\tau J_3(t)dt=-T_h\Delta S+\frac{T_h \mathcal B^{2}}{\tau},
\end{equation}
where $\Delta S\equiv \int_{T_i}^{T_f}(C_V/T)dT$ is the entropy change of the finite-sized cold heat bath. The irreversible entropy production that is inversely proportional to the time duration of the process can be found in several models of heat engines \cite{esp150,izu180,iyy500,sch200} and an experiment \cite{ma210}. 

Using Eqs. (\ref{s}) and (\ref{1d}), we find the total input work in Eq. (\ref{eq.W_def}) as
\begin{equation}\label{1f}
W=E+\frac{T_h \mathcal B^{2}}{\tau},
\end{equation}
where $E$ is the exergy defined as \cite{exe}
\begin{equation}
E\equiv \Delta U-T_h\Delta S.
\end{equation}
The average input power becomes
\begin{equation}\label{1g}
P=\frac{W}{\tau}=\frac{E}{\tau}+\frac{T_h \mathcal B^{2}}{\tau^{2}}.
\end{equation}
In the quasi-static limit, when $\tau\rightarrow \infty$, the input power vanishes as $P\rightarrow 0$~\cite{ron040}. 
Optimizing the performance of refrigerators may 
be dependent on which objective function we adopt
~\cite{yan136,vel324,vel113,de057,all051,de010}, and 
the average input power (Eq. (\ref{1g})) does not have minima.
Therefore, following Ref. \cite{ron040}, we get the process time $\tau$ as
\begin{equation}\label{1h}
\tau=\frac{E}{2P}\left(1\pm\sqrt{1+4T_hP\frac{\mathcal B^{2}}{E^{2}}}\;\right),
\end{equation}
where we adopt only the plus sign as a physically relevant solution.
The COP of the present refrigerator is defined as
\begin{equation}\label{1i}
\epsilon=\frac{Q_c}{W}=\frac{-\Delta U}{E+\frac{T_h \mathcal B^{2}}{\tau}}.
\end{equation}
By substituting the value of $\tau$ into Eq.~(\ref{1i}), we get the inverse of COP as
\begin{equation}\label{1j}
\frac{1}{\epsilon}=\frac{1}{\epsilon_{\rm max}}-\frac{2T_h \mathcal B^{2}P}{\Delta U\left(E+\sqrt{E^{2}+4T_h \mathcal B^{2} P}\;\right)}.
\end{equation}
Here, the maximum COP is given by
\begin{equation}\label{emax}
\epsilon_{\rm max}\equiv \frac{-\Delta U}{E}=\frac{-\Delta U}{\Delta U-T_h\Delta S},
\end{equation}
which serves as the quasi-static counterpart of Carnot COP $\epsilon_{\rm C}$.
This value is recovered for $\Delta U \to T_c\Delta S$ when the cold heat bath has infinite size ($C_V\to \infty$).
In this limit, the finite-time COP in Eq.~(\ref{1i}) also recovers the result with the exergy $E$ being replaced with the quasistatic work $(T_c-T_h)\Delta S$ for infinite-sized heat baths, as consistent with the low-dissipation Carnot refrigerator~\cite{de010,wan2012}.
Moreover, we also recover the consistent result for the efficiency of nonlinear heat engines when they are operated with finite-sized heat baths~\cite{wan012}.

\subsection{Asymmetrical dissipation limits}
Now, we apply the asymmetric dissipation limits to the value of $\gamma$. When $\gamma\rightarrow 1$, we have $\mathcal B\to B_1\equiv \int_{T_i}^{T_f}C_V\sqrt{T_h}/(\sqrt{L_{22}T})dT$. Therefore, the inverse of COP becomes:
\begin{equation}\label{asyl}
\frac{1}{\epsilon}=\frac{1}{\epsilon_{\rm max}}-\frac{2T_h B_1^{2}P}{\Delta U\left(E+\sqrt{E^{2}+4T_h B_1^{2} P}\;\right)}.
\end{equation}
When $\gamma\rightarrow \infty$, we have $\mathcal B\to B_\infty\equiv \int_{T_i}^{T_f}(C_V/\sqrt{L_{22}})dT$. Therefore, we get the inverse of COP as
\begin{equation}\label{asyu}
\frac{1}{\epsilon}=\frac{1}{\epsilon_{\rm max}}-\frac{2T_h B_\infty^{2}P}{\Delta U\left(E+\sqrt{E^{2}+4T_h B_\infty^{2} P}\;\right)}.
\end{equation}
It should be noted that the above result of the case $\gamma=\infty$ was also obtained for a linear irreversible refrigerator working with a finite-sized cold heat bath and an infinite-sized hot heat bath \cite{ron040}.

\subsection{Symmetric dissipation case}
For the symmetric dissipation case $\gamma_c=\gamma_h$, we get $\gamma=2$. Therefore, $\mathcal B\to B_2\equiv \int_{T_i}^{T_f}\frac{C_V}{\sqrt{2L_{22}}}\sqrt{\frac{T_h}{T}+1}\;dT$. The inverse of COP in Eq. (\ref{1j}) becomes
\begin{equation}\label{sym}
\frac{1}{\epsilon}=\frac{1}{\epsilon_{\rm max}}-\frac{2T_h B_2^{2}P}{\Delta U\left(E+\sqrt{E^{2}+4T_h B_2^{2} P}\;\right)}.
\end{equation}
\section{Examples of Finite-sized cold baths}\label{Example}
The temperature dependencies of heat capacity of cold bath $C_V(T)$ and the Onsager coefficient $L_{22}(T)$ are modeled using the well-established asymptotic regimes of three  different physical systems. Namely, the classical ideal gas, the degenerate electron gas, and phonon gas. In each case, for the linear response regime, the Onsager transport coefficient is obtained from the relation $L_{22}(T)=\kappa(T) T^{2}$ \cite{de}. Here, $\kappa$ is the thermal conductivity. Thus, our modeling ensures a consistent description of both thermal bath and heat transport coefficient. These forms are regime-specific rather than universal material laws. These temperature-dependent forms of $C_V$, and $L_{22}$ are implemented in Eqs. (\ref{asyl}), (\ref{asyu}), and (\ref{sym}) in order to study the COP of refrigerator.

\subsection{Ideal gas bath (high-temperature limit)}
For a dilute classical ideal gas in the high-temperature limit, the heat capacity is approximately temperature independent, $C_V=a_I$ \cite{cal}. From the kinetic theory of gases, the thermal conductivity scales as, $\kappa\propto T^{1/2}$ \cite{cha}, which yields $L_{22}=b_I T^{5/2}$. 
\subsection{Degenerate electron gas bath} 
For a metallic bath, the electronic contribution dominates at low temperatures, with $C_V=a_E T$. 
The Wiedemann-Franz law, $\kappa\propto T$ \cite{ash}, gives $L_{22}=b_E T^{3}$. 
\subsection{Phonon gas bath (Debye regime)}
In the low-temperature Debye regime, the phonon contribution leads to $C_V=a_P T^{3}$ \cite{kit}. The thermal conductivity also scales as $\kappa\propto T^{3}$ due to boundary or impurity-limited scattering \cite{zim}, resulting in $L_{22}=b_P T^{5}$.
\subsection{Comparison}
The material-specific prefactors $a_i$, and $b_i$ ($i=I, E, P$) can be restored straightforwardly for quantitative comparison with experiments by using known expressions for the heat capacity and thermal conductivity of the corresponding systems. Thus, while our analysis is presented in a dimensionless form, it remains directly applicable to realistic refrigerator setups through appropriate parameter mapping.
\begin{table}
	\centering
	\begin{tabular}{|c|c|c|c|c|}
		\hline
		\hline
		S. No & System & $B_1(\gamma_c\rightarrow0)$&$B_2(\gamma_c=\gamma_h)$ & $B_{\infty}(\gamma_h\rightarrow0)$\\
		\hline
		\hline
		1&Ideal gas &0.1366 &0.1329  &0.129 \\
		\hline
		2& Electron gas &0.7056 & 0.687 & 0.6677\\
		\hline
		3&  Phonon gas & 6.3246& 6.1637 &5.9969 \\
		
		\hline
		\hline
	\end{tabular}
	\caption{Numerically obtained values of the dissipation parameters, $B_1$, $B_2$, and $B_\infty$ for the three different finite-sized cold baths: ideal gas, electron gas, and phonon gas. Here, we set the initial and final temperatures of the cold heat bath, respectively, as $T_i=T_h=10$, and $T_f=8$. We set $a_I=1$, $a_E=1$, $a_P=1$, $b_I=1$, $b_E=1$, and $b_P=1$ for simplicity.}
	\label{table:pe}
\end{table}

\begin{figure*}[hpt]
\centering
\includegraphics[scale=0.85,angle=0]{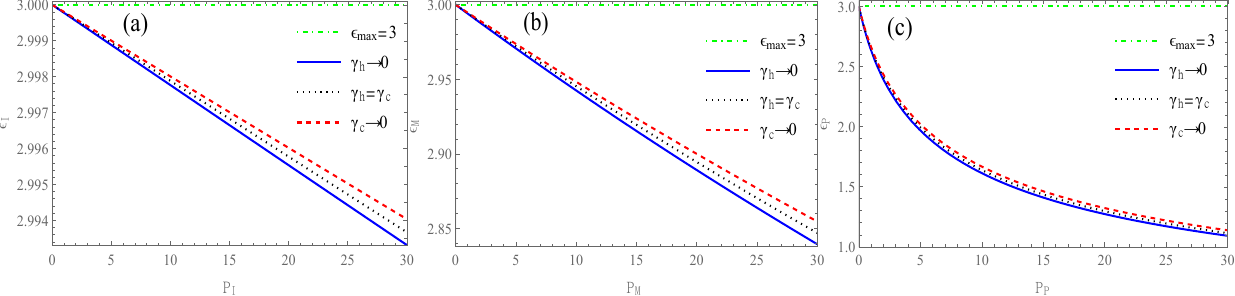}
\vspace{-0.2cm}
\caption{\label{fig:figcop}The COP of the refrigerator is plotted as a function of average input power. For an ideal gas cold heaet bath (a), for degenerate electronic gas cold heat bath (b), and for phonon gas cold bath at low-temperature (c). The green dot-dashed line indicates $\epsilon_{\rm max}=3$. The solid blue curve represents the lower asymmetric dissipation limit, $\gamma_h\rightarrow0$ ($\gamma\rightarrow1$) (Eq.~(\ref{asyl})). The dotted black curve represents the symmetric dissipation case $\gamma_h=\gamma_c$ ($\gamma$=2) (Eq.~(\ref{sym})). The dashed red curve represents the upper asymmetric dissipation limit, $\gamma_c\rightarrow0$ ($\gamma\rightarrow\infty$) (Eq.~(\ref{asyu})). We set following values, $E=50$, and $-\Delta U=\epsilon_{\rm max} E$ (from Eq. (\ref{emax})).}
\end{figure*}

In Fig.~\ref{fig:figcop}, we plot the COP obtained from Eqs.~(\ref{asyl}), (\ref{asyu}), and (\ref{sym}) as a function of the average input power for three distinct finite-sized cold baths: (a) ideal gas, (b) degenerate electron gas, and (c) phonon gas. 
See Table~\ref{table:pe} for $B_1$, $B_2$, and $B_\infty$ of each heat bath model.
For the parameter values considered in this study, the COP decreases approximately linearly with average input power for both ideal gas and electron gas baths. However, nonlinear decrease is observed for the phonon bath. Notably, the decrease of COP with increasing input power is significantly stronger in the phonon bath as compared to the other two baths. This enhanced sensitivity of the phonon bath due to the strong temperature dependence of the Onsager coefficient ($L_{22}\propto T^{5}$) may lead to a rapid increase in dissipation under finite-time operation. The physical reasoning of the COP decrease with input power can be understood from the interplay between operation time $\tau$ and irreversibility. As the input power increases, the process time $\tau$ decreases (i.e., decrease in the cooling process time) (see Eq. (\ref{1h})). This decrease in cycle time enhances irreversible entropy production, which in turn lowers the COP \cite{ron040} (see Eq. (\ref{1f})). The larger decrease of COP observed in the phonon bath originates from its steeper temperature dependence of transport coefficient, which amplifies the dissipation under fast driving. Furthermore, in the asymmetric dissipation limit $\gamma_c\rightarrow0$, the COP reaches its maximum value as compared to both the symmetric case and the other asymmetric limit. This trend is consistently observed in all three finite-sized cold heat bath models considered in this work.

\begin{figure*}[hpt]
	\centering
	\includegraphics[scale=0.85,angle=0]{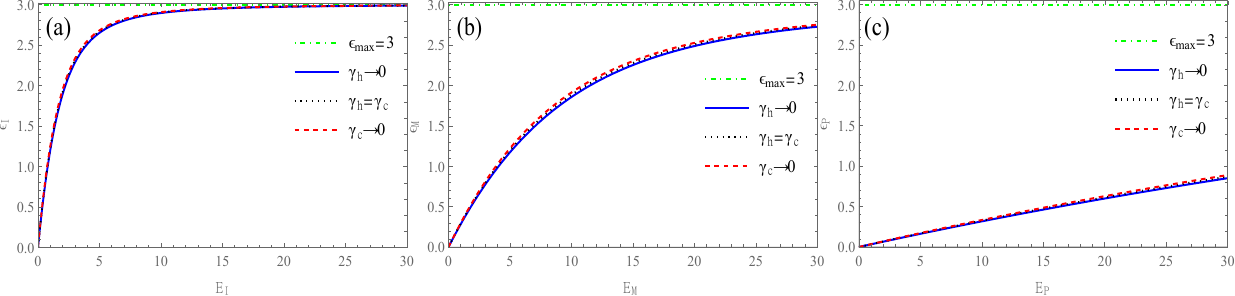}
	\vspace{-0.2cm}
	\caption{\label{fig:figcopex} The COP of the refrigerator is plotted as a function of exergy. For an ideal gas cold heat bath (a), for degenerate electronic gas cold bath (b), and for phonon gas cold heat bath at low-temperature (c). The green dot-dashed line indicates $\epsilon_{\rm max}=3$. The solid blue curve represents the lower asymmetrical dissipation limit, $\gamma_h\rightarrow0$ ($\gamma\rightarrow1$) (Eq.~(\ref{asyl})). The dotted black curve represents the symmetric dissipation case, $\gamma_h=\gamma_c$ ($\gamma=2$) (Eq.~(\ref{sym})). The dashed red curve represents the upper asymmetrical dissipation limit, $\gamma_c\rightarrow0$ ($\gamma\rightarrow\infty$) (Eq.~(\ref{asyu})). We set the following values, $P=20$, and $-\Delta U=\epsilon_{\rm max} E$.}
\end{figure*}

In Fig. \ref{fig:figcopex}, we plot the COP obtained from Eqs.~(\ref{asyl}), (\ref{asyu}), and (\ref{sym}) as a function of exergy $E$ for a fixed average input power. We find that the COP increases nonlinearly with exergy for both the ideal gas and degenerate electron gas cold baths, whereas a nearly linear dependence is observed for the phonon gas bath. In particular, the rate of increase of COP with exergy is significantly low in the phonon bath case as compared to the other two cold baths. This behavior can be explained from the relation between the exergy and the process time and associated dissipation. As the exergy $E$ increases, the cycle time $\tau$ increases (see Eq.~(\ref{1h})), effectively allowing the system to operate closer to the quasi-static regime. Consequently, irreversible entropy production is reduced, leading to the enhancement of the COP. In the limiting case $E\rightarrow\infty$, the COP approaches its maximum value $\epsilon_{\rm max}$ (see Eq. (\ref{1j})), which is consistent with the reversible case. The weaker enhancement observed for the phonon bath originates from its stronger temperature dependence of the transport coefficients, which limits the reduction of dissipation even as the cycle time increases. As a result, the gain in performance with exergy remains comparatively low. Finally, for the parameter regime considered here, we do not observe any significant variation in the COP under different asymmetric dissipation limits, indicating that the exergy-controlled behavior dominates over dissipation asymmetry in this case. This indicates that, in contrast to power-controlled optimization, the exergy primarily governs the proximity to reversibility, rendering dissipation asymmetry subdominant in determining the COP.

\section{Conclusions}\label{Conclusion}
In this work, we have studied the performance of minimally nonlinear irreversible refrigerators operating between a finite-sized cold heat bath and an infinite-sized hot heat bath (environment). We assume that initially the temperature of the finite-sized cold heat bath is equal to or smaller than that of the environment. The refrigerator starts to transport heat from a finite-sized cold heat bath to a hot environment by consuming the input power. Using the minimally nonlinear irreversible model and the Euler-Lagrange equation, we find the optimal temperature variation (see Eq. (\ref{1b})), which minimizes the input work. We obtained the COP as a function of average input power and exergy. When we apply the asymmetric dissipation limits, we get the lower and upper bounds of the COP (see Eqs. (\ref{asyl}) and (\ref{asyu})). Further, we also find the COP in the symmetric dissipation case (see Eq. (\ref{sym})).

To elucidate the impact of realistic finite-sized cold heat baths on the coefficient of performance, we considered three representative scenarios: an ideal gas, a degenerate electron gas, and a low-temperature phonon gas. For the parameter regime explored here, we find that the COP decreases approximately linearly with the average input power for both the ideal gas and electron gas baths. In contrast, for the phonon bath in the low-temperature (Debye) regime, the COP decreases  nonlinearly with input power (see Fig. \ref{fig:figcop}). We further analyzed the dependence of COP on the exergy. We find that, for the ideal gas and electron gas baths, the COP increases nonlinearly with exergy. By contrast, the phonon bath displays an approximately linear increase of COP with exergy (see Fig.~ \ref{fig:figcopex}). Notably, the COP in the phonon case is significantly more sensitive to variations in input power while exhibiting a comparatively weaker dependence on exergy (see Figs.~\ref{fig:figcop}(c) and \ref{fig:figcopex}(c)).

These results highlight that the nature of the finite-sized bath—through its temperature-dependent thermodynamic and transport properties—plays a crucial role in determining the COP of the refrigerators. In particular, the strong temperature dependence associated with the phonon bath leads to enhanced dissipation under finite-time operation, thereby amplifying the sensitivity of COP to input power while limiting its improvement with increasing exergy.

Since our framework explicitly incorporates both the finite-size effects of the cold bath and dissipation by using a minimally nonlinear irreversible refrigerator, it provides direct insights into the design and operation of realistic refrigeration systems. In particular, the COP can be enhanced by reducing the dissipation strength associated with the cold bath, $\gamma_c$ as given in Eq. (\ref{asyu}) and Fig.~\ref{fig:figcop}. This regime corresponds to the upper asymmetric dissipation limit, where irreversible losses to the cold bath are minimized. Practically, such a condition can be approached by appropriately tuning the system parameters to satisfy $T_h/L_{11}\approx\gamma_h$ (see Eq.~(\ref{h})), thereby effectively redistributing the dissipation toward the hot bath and improving the COP. This highlights that controlling the distribution of dissipation between the baths is a key strategy for optimizing finite-time refrigeration performance.

It will be interesting to study the performance of a thermoelectric refrigerator with a finite-sized cold heat bath and an infinite-sized hot heat bath with and without breaking the time-reversal symmetry of the Onsager coefficients \cite{ben230,bra070} beyond the present setup. It will also be challenging to study how finite-sized heat baths could affect the performance of Brownian refrigerator~\cite{van210}.

\section*{ACKNOWLEDGMENTS}
I.I would like to thank The University of Tokyo for the hospitality during the visit when the initial part of the project was discussed. We thank Zhen Li for translating Ref. \cite{ron040} in Chinese to English. I.I is also thankful to Shrihari Gopalakrishna and Pinaki Swain for their valuable comments and suggestions. The authors made limited use of Large Language Models (LLMs) to improve the clarity to part of the manuscript.
\linebreak

I.I and Y.I contributed equally to this work.

\appendix
\section{Justification for minus sign}\label{appendix_root1}
\begin{figure}[h!]
	\centering
	\includegraphics[scale=0.233]{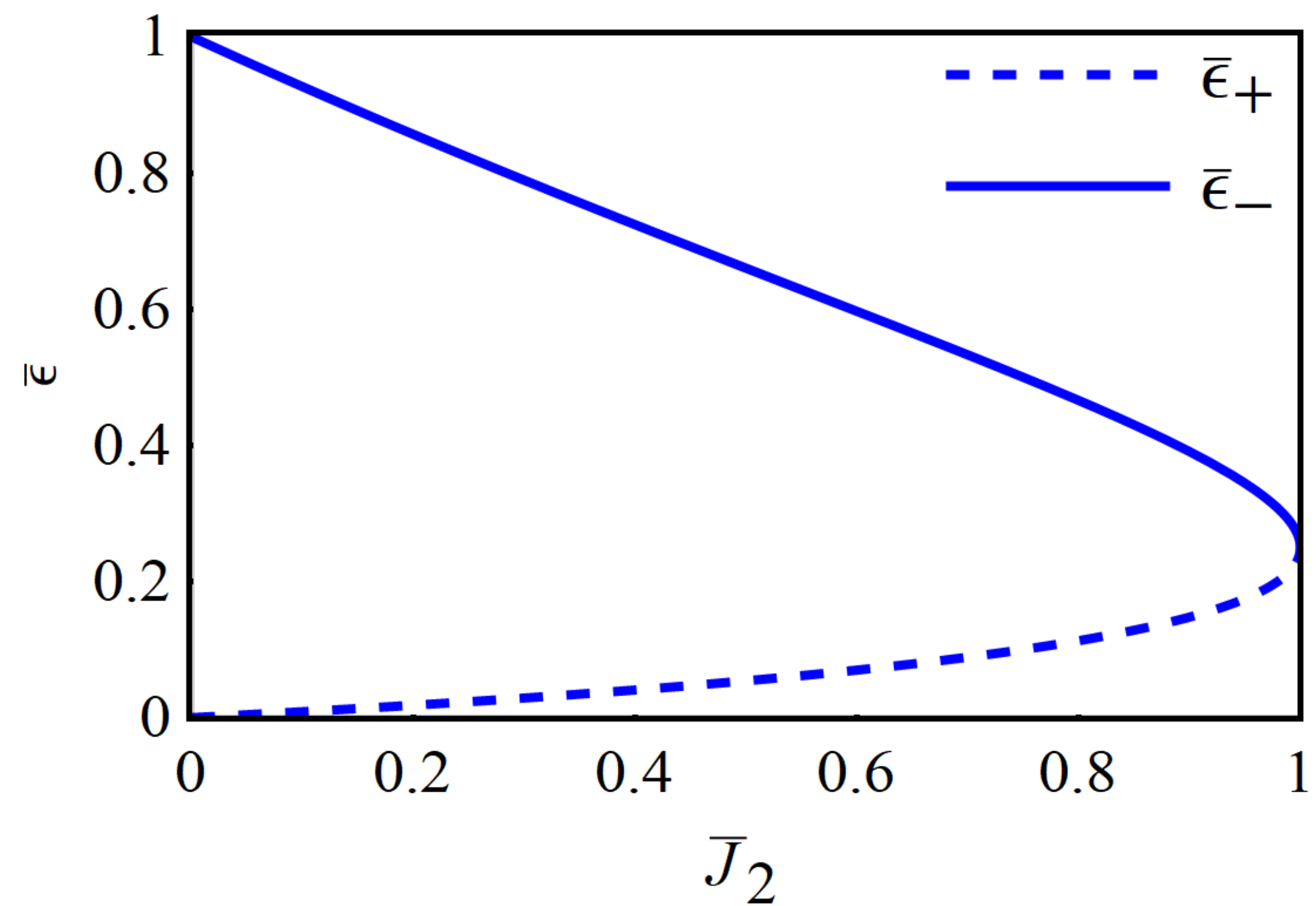}
	\caption{The normalized COP is plotted as a function of the normalized cooling power. The dashed and solid curves represent the $\bar{\epsilon}_+$ (positive root), and $\bar{\epsilon}_-$ (negative root) of Eq.~(\ref{b6}), respectively. Here, we set $\epsilon_{\rm C}^T=1$.}
	\label{fig.cop_nor}
\end{figure}

We show that it is sufficient to consider only the negative root case in Eq.~(\ref{k1}) (or Eq.~(\ref{j})) for optimization. 
For simplicity, we consider the heat flux absorbed from the cold heat bath (Eq.~(\ref{e})) with the tight-coupling condition
\begin{equation}
J_2=\frac{L_{21}}{L_{11}}J_1-\gamma_cJ_1^{2}.
\end{equation} 
Maximizing the above cooling power with respect to $J_1$, we get $J_1^{\star}=L_{21}/(2L_{11}\gamma_c)$. Hence, the maximum cooling power becomes 
\begin{equation}
J_2^{\star}=\frac{L_{21}^{2}}{4L_{11}^{2}\gamma_c}.
\end{equation} 
The normalized cooling power can be defined as
\begin{equation}
\bar{J_2}=\frac{J_2}{J_2^{\star}}=\frac{4L_{11}\gamma_c}{L_{21}}J_1-\frac{4L_{11}^{2}\gamma_c^{2}}{L_{21}^{2}}J_1^{2}.
\end{equation}
Solving the above equation for $J_1$, we get
\begin{equation}\label{b4}
J_1^{\pm}=\frac{L_{21}}{2L_{11}\gamma_c}\left(1\pm\sqrt{1-\bar{J}_2}\right).
\end{equation}
The COP reads
\begin{equation}\label{b5}
\epsilon=\frac{J_2}{P}=\frac{\frac{L_{21}}{L_{11}}J_1-\gamma_cJ_1^{2}}{\frac{L_{12}}{L_{11}\epsilon_{\rm C}^T}J_1+\frac{T_h}{L_{11}}J_1^{2}}=\frac{\frac{L_{21}}{L_{11}}-\gamma_cJ_1}{\frac{L_{12}}{L_{11}\epsilon_{\rm C}^T}+\frac{T_h}{L_{11}}J_1}.
\end{equation}
where $\epsilon_{\rm C}^T \equiv T/(T_h-T)$.
Substituting Eq.~(\ref{b4}) into Eq.~(\ref{b5}), we get the normalized COP as
\begin{equation}\label{b6}
\bar{\epsilon}_{\pm}\equiv\frac{\epsilon}{\epsilon_{\rm C}^T}=\frac{1\mp\sqrt{1-\bar{J}_2}}{2+\epsilon_{\rm C}^T(1+\frac{\gamma_h}{\gamma_c})(1\pm \sqrt{1-\bar{J}_2})}.
\end{equation}
To identify the optimal operating regime, we consider the symmetric case ($\gamma_c=\gamma_h$) as an example shown in Fig.~\ref{fig.cop_nor}. The comparison between the two solution branches reveals that the negative-root branch yields higher performance. Therefore, it is more suitable for practical operation than the positive-root branch. This justifies our analysis only using the negative-root solution. Plots analogous to those shown in Fig. \ref{fig.cop_nor} were reported for linear irreversible thermoelectric heat engines in Ref. \cite{ben1}.

\section{Comparison between numerical and analytical solutions of $T(t)$}\label{comparison}

\begin{figure}[t!]
\centering
\includegraphics[scale=0.73]{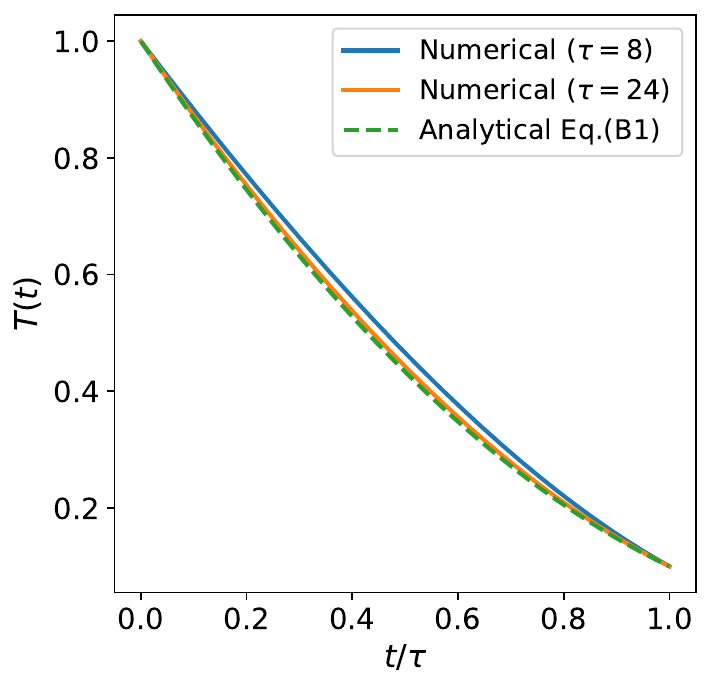}
\caption{Comparison between numerical and analytical results of $T(t)$ for $\tau=8$ and $\tau=24$. We used $T_i=T_h=1$ and $T_f=0.1$.}
\label{fig.EL}
\end{figure}

We compare the solutions of the Euler-Lagrange equation~\eqref{m} when using Eq.~\eqref{k1} and when using Eq.~\eqref{w} as $J_3$.
Here, we adopt $\gamma=L_{22}=C_V=1$ as a simple case.
When we use Eq.~\eqref{k1} as $J_3$, we numerically solve it by using a shooting method with the boundary conditions $T(0)=T_i$ and $T(\tau)=T_f$.
When we use Eq.~\eqref{w} as $J_3$, we can obtain its solution explicitly as
\begin{eqnarray}\label{eq.T_analytic}
T(t)=\left(\sqrt{T_i}+\frac{\sqrt{T_f}-\sqrt{T_i}}{\tau}t\right)^2,
\end{eqnarray}
by solving Eq.~\eqref{1b}.

In Fig.~\ref{fig.EL}, we compared the numerical solution and the analytical solution Eq.~\eqref{eq.T_analytic}. We find that when $\tau$ is sufficiently large, the numerical result and Eq.~\eqref{eq.T_analytic} show a good agreement. This is consistent with the condition Eq.~\eqref{eq.approx_cond} that requires sufficiently small changing rate of the temperature.


\begin{thebibliography}{100}
\bibitem{exe}
It has to be noted that the exergy is defined for a heat engine as the maximum available work that can be extracted from a finite-sized hot heat source until it reaches thermal equilibrium with the infinite-sized cold bath \cite{mora}. However, for refrigerators hot and cold baths never
reach thermal equilibrium. With a little abuse of terminology, we use exergy as a minimum work required to cool the finite-sized cold bath from an initial temperature $T_i$ to the final temperature $T_f < T_i$.
\end{thebibliography}
\end{document}